\documentstyle[12pt,aaspp4]{article}

\slugcomment{To Appear in the {\em Astrophysical Journal Letters}}
\begin{document}
\def\kmsmpc{${\rm km\,s^{-1}\,Mpc^{-1}}$} 
\def\qo{${\rm q_{o}}$}
\def\Ho{${\rm H_{o}}$}                          

\title {\bf The FIRST Radio-Loud Broad Absorption Line QSO and
Evidence for a Hidden Population of Quasars}

\author{Robert H. Becker}
\affil{University of California at Davis\\
and\\
Institute for Geophysics and Planetary Physics\\
Lawrence Livermore National Laboratory\\
bob@igpp.llnl.gov}

\author{Michael~D.~Gregg}
\affil{Institute for Geophysics and Planetary Physics\\
Lawrence Livermore National Laboratory\\
gregg@igpp.llnl.gov}

\author{Isobel M. Hook}
\affil{European Southern Observatory\\
imh@eso.org}

\author{Richard G.~McMahon}
\affil{Institute of Astronomy, Cambridge\\
rgm@ast.cam.ac.uk}

\author{Richard L. White}
\affil{Space Telescope Science Institute\\
rlw@stsci.edu}

\author{David J. Helfand}
\affil{Columbia Astrophysics Laboratory\\
djh@astro.columbia.edu}

\begin{abstract}

We have discovered two low-ionization broad absorption line quasars in
programs to obtain optical spectra for radio-selected quasar
candidates from the VLA FIRST Survey (Becker, White, \& Helfand 1995).
Both belong to the extremely rare class of BAL QSOs that exhibit
narrow absorption lines from metastable excited levels of Fe~II and
Fe~III.  Until now, there was just a single object in this class,
0059-2735 (Hazard et al.\ 1987).  In addition, one of our new objects
is the {\em first known radio-loud BAL QSO.}  The properties of these
three unusual objects suggest a trend of increasing radio luminosity
with the amount of absorption to the quasar, and are perhaps
transition objects between radio-loud and radio-quiet quasars.

The two new objects are from a radio-selected sample comprising less
than 200 quasars; one is heavily attenuated at optical wavelengths in
the observed frame.  These objects would be easily overlooked by most
optical QSO searches; their abundance in the radio sample suggests
that they may be representatives of a largely undetected component of
the quasar population, perhaps as numerous as ordinary low-ionization
BAL~QSOs which constitute 1-2\% of all QSOs.  

\end{abstract}


\section {Introduction}
     
Radio-loud quasars and broad absorption line (BAL) quasars (QSOs) each
constitute $\sim 10\%$ of the total quasar population (Foltz et al.\
1990; Weymann et al.\ 1991) but have always appeared to be mutually
exclusive classes (Stocke et al.\ 1992).  There is little agreement
about what determines quasar radio luminosities or what the
relationship is between broad absorption lines and radio power.  The
BAL~QSOs can be further subdivided into high and low-ionization
objects (Hi-BALs and Lo-BALs).  The Lo-BALs make up only 10\% of the
total BAL QSO population (Weymann et al.\ 1991) and are usually
recognized by the presence of broad absorption from Si~II, Mg~II, and
Al~III.  In the Lo-BAL population, there has for a decade been only one
object, 0059-2735, which also shows narrow absorption lines from
metastable excited states of Fe~II and Fe~III (Hazard et al.\ 1987).  We
report here the discovery of two more ``iron Lo-BALs''. 

We have undertaken two programs to identify candidate quasars drawn
from the NRAO\footnote{The National Radio Astronomy Observatory is
operated by Associated Universities, Inc., under cooperative agreement
with the National Science Foundation} Very Large Array (VLA)
FIRST\footnote{The FIRST Survey World Wide Web homepage is
http://sundog.stsci.edu} Survey (Becker, White, \& Helfand 1995; White
et al.\ 1997).  Both programs select optical samples for spectroscopy
by matching FIRST radio sources to stellar counterparts within
$1\farcs2$ from the APM POSS I catalog (McMahon \& Irwin 1992).  The
FIRST Bright Quasar Survey (FBQS, Gregg et al.\ 1996) is developing a
new, complete sample of radio-selected quasars brighter than 17.5
magnitude on the E plates and with $O - E < 2.0$\footnote{The POSS~I O
and E bands roughly correspond to the more familiar Cousins B and R.
The APM magnitudes have an accuracy of $\sim 0.4$ magnitudes, though
the colors are probably more accurate (McMahon \& Irwin 1992).}.  The
second program is searching specifically for high redshift quasars
among very red optical counterparts with $O-E > 1.0$ with the
additional constraint that the sources also have flat radio spectra
between 6 and 20 cm (Hook et al.\ 1996).  The FIRST Survey quasar
search programs are unique in that they go to much fainter radio
levels than previous radio-selected QSO surveys: the 1 mJy flux limit
picks up radio-quiet objects out to $z \approx 2$.  Roughly half of
the objects in the FBQS pilot sample are, in fact, radio-quiet (Gregg
et al.\ 1996).

For the past two years we have been taking 5\AA\ resolution spectra of
these quasar candidates using the Lick 3m Shane telescope, the Kitt
Peak National Observatory\footnote{Kitt Peak National Observatory,
NOAO, is operated by the Association of Universities for Research in
Astronomy, Inc. (AURA), under cooperative agreement with the National
Science Foundation} 2.1m and 4m telescopes, and the La Palma 2.5m
Isaac Newton telescope.  To date, 136 new quasars with R $\lesssim
17.5$ have been found in the FBQS and $\sim 30$ additional quasars
have been found in the high redshift program.

Among the new QSOs in the FIRST Survey samples, there are 4 BAL~QSOs.
Two are of the common Hi-BAL variety, but the other two, one from each
program, have very unusual spectra, placing them in the iron Lo-BAL
class with the heretofor unique object 0059-2735 (Hazard et al.\
1987).  Additionally, one of these two is the first known radio-loud
BAL~QSO.

\section {Observed Properties of 0840+3633 and 1556+3517}

\subsection {Optical Spectra and Photometry}

Optical spectra obtained with the Lick Observatory 3m telescope for
FIRST J084044.5+363328 (0840+3633) and FIRST J155633.8+351758
(1556+3517) are displayed in Figure~1, shifted to the rest frame,
along with the spectrum of 0059-2735 (from Weymann et al.\ 1991).
Unlike most quasar spectra, which are characterized by strong, broad
emission lines, these BAL~QSO spectra are completely dominated by
their absorption features; emission lines are relatively weak or
absent.  In Figure~1, the distinguishing absorption features are
indicated; an asterisk denotes a metastable iron feature.  In the
spectrum of 0840+3633, the unabsorbed remnant of the Mg~II 2800\AA\
emission line yields a redshift of 1.22.  Its spectral features
are very similar to those of 0059-2735, albeit with deeper absorption
troughs, and it is unmistakably a more extreme example of this class
of BAL~QSO.  The spectrum of 1556+3517 has no prominent emission
lines, but the absorption features are so similar to 0840+3633 and
0059-2735 that there is again no doubt that it is the same kind of
object with even higher absorption.  Not only are the broad absorption
lines nearly saturated, but the continuum is heavily absorbed.  The
absorption features indicate a redshift of 1.48.  In the two new BALs,
the broad lines of Al~II, Al~III, and Mg~II have nearly complete
absorption in their cores.  The absorption due to metastable excited
lines of Fe~II and Fe~III (Hazard et al.\ 1987) is also deeper in the
two new objects than in 0059-2735.

For 0840+3633, the APM POSS~I magnitudes are O = 17.3 and E = 15.9,
while for 1556+3517, O = 21.2 and E = 18.7.  Both objects
are redder than typical intermediate redshift quasars, for which $O -
E \approx 0.5$ (FBQS).  For 0059-2735, Hazard et al.\ (1987) estimate
R = 17.1 and Hewett et al.\ (LBQS, 1995) list $B_J = 18.13$.  We adopt
\Ho = 50 \kmsmpc, \qo = 0.1, and an optical spectral index
$\alpha_{\rm opt} = -1.0$ to compute the absolute R-band magnitudes. 
The pertinent data for the three iron Lo-BALs are listed in Table~1.

\subsection {Radio Properties}

Both new BAL~QSOs are point sources in the FIRST catalog.  The 1.4~GHz
flux densities for 0840+3633 and 1556+3517 are 1.3 and 30~mJy,
respectively.  The Greenbank 5~GHz catalog (Becker, White, \& Edwards
1991) lists a flux density of 27~mJy for 1556+3517, giving a spectral
index of -0.1, assuming no variability.  Stocke et al.\ (1992) provide
an upper limit of 0.36 mJy for the 5~GHz flux density of 0059-2735.
The flat radio spectrum for 1556+3517 could be indicative of
relativistic beaming, making a calculation of radio luminosity
problematic, but we have computed L$_{1400}$ for all three sources
assuming isotropic radiation (Table~1) and adopting a spectral index
of -0.1.

We have computed the ratio of 5~GHz radio to 2500\AA\ optical flux,
log(R), using equations 1-3 of Stocke et al.\ (1992).  No
K-corrections have been applied.  The BAL 0059-2735 is radio-quiet
with log(R)$\approx 0$, 0840+3633 is ``radio-moderate'' with log(R) =
0.4, 1556+3517 has log(R)$ > 3$.  The usual dividing line between
radio-loud and radio-quiet is log(R) $\approx 1$ (e.g.\ Stocke et
al.), so, even with the uncertainties in the optical photometry, {\em
1556+3517 is the first known radio-loud BAL~QSO.}

Could the conclusion that 1556+3517 is radio-loud be a consequence of
its large optical absorption?  Its absolute B magnitude is $\sim 3.4$
magnitudes fainter than 0059-2735 and 0840+3633; if this difference is
attributed entirely to attenuation, then log(R) drops to $\sim 1.8$,
still a radio-loud object.  Additionally, it has ${\rm L_{1400}=
10^{33.5}~ergs~s^{-1}Hz^{-1}}$, well above the usual divide between
radio-loud and radio-quiet of $10^{32.5}$ based on radio power alone
(Schneider et al.\ 1992; Stocke et al.\ 1992).

\section {Discussion}

\subsection{Why Does FIRST Find Iron Lo-BALs?}

Francis, Hooper, and Impey (1993) have shown that BAL~QSOs tend to be
found closer to the radio-loud/radio-quiet divide than typical
optically selected QSOs.  They estimated that BAL~QSOs are
overabundant by a factor of 10 in the ``radio-moderate'' ($0.2 <
log(R) < 1.0$) population compared to the radio-quiet population, and
one of the brightest radio-moderates in their sample is a Lo-BAL,
1235+1807B.  The low radio flux limit of the FIRST Survey leaves it
sensitive to radio-quiet and radio-moderate objects with $z \lesssim
2$, so the presence of ordinary Lo-BALs in the FIRST quasar samples
would not be surprising.  Out of 170 newly identified quasars in the
FIRST Survey searches, only two Lo-BALs have been found so far and
both have highly unusual spectra placing them in the rare iron Lo-BAL
class, defined by 0059-2735 (Hazard et al.\ 1987).  In the current
census of $\sim 8600$ quasars of V\'{e}ron-Cetty \& V\'{e}ron (1996),
0059-2735 is the only other such recognized object.  The FIRST Survey
is apparently more effective at finding these unusual quasars than
previous surveys.

The rarity of iron Lo-BALs up until now can to some extent be
attributed to the absence of UV excess or strong emission lines in
their optical spectra as well as their relatively low radio flux
densities.  Even multicolor high-$z$ quasar surveys (e.g., Warren,
Hewett, and Osmer 1994) have not detected this class of quasar.
Although Hazard et al.\ (1987) point out that the appearance of
0059-2735 in their discovery objective prism data drew attention
because of its strong Al~II and Al~III absorption, objects as heavily
attenuated as 1556+3517 probably masquerade effectively as M~stars in
such surveys.  The FBQS has a relatively weak color selection,
including objects as red as $O-E = 2$, while the high redshift survey
intentionally targets even redder stellar objects, and neither rely on
objective prism data, so objects like 0840+3633 and 1556+3517 are not
selected against in the optical.

The spectra of the three iron Lo-BALs (Figure~1) suggest a correlation
between column density of low-ionization clouds and radio emission.
Based on the Fe~II absorption lines and continuum attenuation,
0059-2735 has the lowest column density, followed by 0840+3633, and
then 1556+3517; ranking by radio luminosity yields the same order.
The severe attenuation of the continuum shape of 1556+3517 also
suggests the presence of large amounts of dust.  Although derived from
a very small sample, this correlation suggests that high column
densities in these BAL~QSOs are associated with stronger radio
emission.  The preponderance of known BAL~QSOs have much lower column
densities of low-ionization clouds, consistent with the relatively low
radio power of all BAL~QSOs discovered to date.  If the correlation
between extinction in the optical and radio luminosity is correct,
then as the iron Lo-BALs become harder to detect in the optical, they
become more prominent in the radio and will naturally be more
plentiful in radio-selected samples.  Based purely on these empirical
results, we propose that most BAL~QSOs are radio-quiet and will not
show up in a radio-selected sample, while the iron Lo-BALs are a
special subpopulation which occupy the transition region between
radio~quiet and radio~loud objects.  The discovery of two iron Lo-BALs
out of $ < 200$ quasars suggests that they make up $\sim 1-2\%$ of the
FIRST quasars, comparable to the rate of Lo-BALs in the general quasar
population as currently understood.

A possible additional high-redshift (z = 2.33) member of the iron
Lo-BALs is Hawaii~167, analyzed by Cowie et al.\ (1994) and Egami et
al.\ (1996).  The spectrum of Hawaii~167 (Figure~2 of Cowie et al.)
does show marked similarities with 0059-2735 but lacks strong
Mg~II~2800 emission and does not have particularly broad absorption
troughs.  The discovery of the two new iron Lo-BAL objects eases the
transition between 0059-2735 and Hawaii~167; 0840+3517 is intermediate
while 1556+3517 has even less Mg~II emission and a more attenuated
continuum than Hawaii~167, though its absorption features appear
broader.  If our suggestion of a link between radio power and column
density of absorbing material is correct, then Hawaii~167 should have
a radio luminosity between that of 1556+3517 and 0840+3633.
Hawaii~167 has not been detected in the 20cm VLA NVSS Survey (Condon
et al.\ 1996), implying an upper limit of $\sim 2.5$ mJy and
log(L$_{1400}) < 32.8$, leaving room for it to fit the trend.

Hawaii~167 is quite faint in the optical, B = 23.0 (Cowie et al.).  It
was found in a complete spectroscopic survey of IR-selected objects in
a region covering only 77 arcmin$^2$.  Cowie et al.\ (1994) argue that
finding such an object in a small area survey indicates that these
objects may be quite common, albeit at faint optical magnitudes,
consistent with the results of the FIRST quasar samples.

\subsection{Dusty QSO Interpretation}

Voit, Weymann, \& Korista (1993) have proposed a model for Lo-BALs in
which a nascent QSO is embedded in a region with a high rate of
massive star formation, enshrouded with dust.  In their scenario,
these objects then evolve into more ordinary Hi-BALs with time, as
their dust shrouds dissipate.  If indeed 1556+3517 were to evolve into
a more ordinary BAL, its radio luminosity would have to diminish.
Egami et al.\ (1996) extend this model to Hawaii~167 and 0059-2735,
explaining the iron Lo-BALs as a combination of heavily reddened QSO +
starburst.  They point out that such objects would likely be missed by
optical surveys because the high dust content suppresses the rest
frame UV (observed optical) and because these objects lack strong
emission lines.  They further note that if the starburst activity has
not begun, these objects will be ``completely dark'' in the optical,
implying that there may be a large, undetected population of these
hybrid starburst/quasar objects.

This model helps to explain the FIRST Survey sensitivity to iron
Lo-BALs: our radio-selected samples are at least partly immune to the
optical selection effects that may cause these objects to be
overlooked.  If so, the FIRST Survey is now picking up the brightest,
and possibly lowest redshift, tip of the iceberg.  These objects have
not yet acquired strong emission lines, but have evolved to the point
where they can be seen through the surrounding dust.  Additional
arguments for populations of undetected, dusty quasars have been put
forth by Low et al.\ (1989), Sprayberry \& Foltz (1992), and Webster
et al.\ (1995).  Just how numerous the iron Lo-BALs are will be
determined as the FIRST QSO surveys progress and IR surveys become
more common.

\section{Conclusions}

We have discovered two low-ionization BAL~QSOs with strong absorption
from metastable excited states of Fe~II and III (``iron Lo-BALs'').
There has been just a single member of this class until now,
0059-2735, with Hawaii~167 another likely example.  One of the new
objects, 1556+3517, is the first known radio-loud BAL~QSO.  The three
iron Lo-BALs suggest a correlation between radio luminosity and
absorption.  We suggest that the iron Lo-BALs are transition objects
evolving from radio-loud to radio-quiet BAL systems as the QSO emerges
from the enshrouding material.  The two new FIRST Survey objects and
Hawaii~167 were all discovered in limited surveys at wavelengths
longer than optical, suggesting that these objects escape detection in
most optical surveys and may be common, perhaps equal in population to
the radio-quiet Lo-BALs.

Whether considered extreme or transition, the iron Lo-BALs may hold
clues to several aspects of the quasar phenomenon, starburst activity
in the early Universe, and the relation of both to galaxy formation.
Egami et al.\ (1996) emphasize that the heavy obscuration of the
central QSO allows the surrounding galaxy starlight to be detected,
opening a window on QSO galaxy hosts at higher redshift.  If this is
accurate, then 1556+3517 should be even more galaxy-like in the IR
than Hawaii~167 because of its higher extinction of the QSO source.
It is possible that these objects are a ``missing link'' between
galaxies and quasars.

\acknowledgments

We acknowledge Jules Halpern for first recognizing the nature of
0840+3633 and for additional helpful discussions.  We thank Paul
Hewett for the spectrum of 0059-2735 and Michael Brotherton for many
useful comments and suggestions.  We acknowledge support from the
NRAO, the NSF (grants AST-94-19906 and AST-94-21178), the IGPP/LLNL
(DOE contract W-7405-ENG-48), the STScI, the National Geographic
Society (grant NGS~No.\ 5393-094), NATO (grant CRG~950765), and Sun
Microsystems.  This paper is Contribution Number 620 of the Columbia
Astrophysics Laboratory.

\pagebreak

\pagebreak


\begin{deluxetable}{lcccccccccc}
\footnotesize
\tablenum{1}
\tablewidth{0pt}
\tablecaption{BAL Quasar Vital Statistics\tablenotemark{a}}
\tablehead {
\colhead {ID} &
\colhead {RA} &
\colhead {Dec} &
\colhead {z} &
\colhead {B} &
\colhead {R} &
\colhead {M$_{\rm R}$} &
\colhead {S$_{1400}$\tablenotemark{b}} &
\colhead {S$_{5000}$} &
\colhead {Log(L$_{1400}$)} &
\colhead {Log(R$^{*})$} \\
\colhead {} &
\multicolumn {2}{c}{(J2000)} &
\colhead {} &
\colhead {} &
\colhead {} &
\colhead {} &
\colhead {(mJy)} &
\colhead {(mJy)} &
\colhead {(ergs s$^{-1}$ Hz$^{-1}$)} &
\colhead {}
}
\startdata
$0059-2735$ & 01 02 17.1 & $-27$ 19 51
& 1.62 & 18.1 & 17.1 & $-28.9$ 
& \nodata  & $< 0.36$\tablenotemark{c} 
& $<31.71$\tablenotemark{e}  & $< 0.04$ \nl
$0840+3633$ & 08 40 44.5 & $+36$ 33 28
& 1.22 & 17.3 & 15.9 & $-29.3$ 
& \phn1.6 & \nodata                  
& $\phm{<\ }32.03$ & $\phm{<\ }0.37$ \nl
$1556+3517$ & 15 56 33.8 & $+35$ 17 58 
& 1.48 & 21.2 & 18.7 & $-27.0$ 
& 30.6 & \phm{<}27.0\tablenotemark{d}    
& $\phm{<\ }33.50$ & $\phm{<\ }3.18$ \nl
\enddata
\tablenotetext{a}{for H$_{\rm o} = 50$, q$_{\rm o} = 0.1$,
$\alpha_{rad}=-0.1$, $\alpha_{opt}=-1.0$}
\tablenotetext{b}{FIRST Survey, Becker et al.~1995}
\tablenotetext{c}{Stocke et al.~1992}
\tablenotetext{d}{GB 5GHz Survey, Becker et al.~1991}
\tablenotetext{e}{Estimated from S$_{5000}$ upper limit}
\end{deluxetable}


\begin{figure}[h]
\figurenum{1}
\plotone{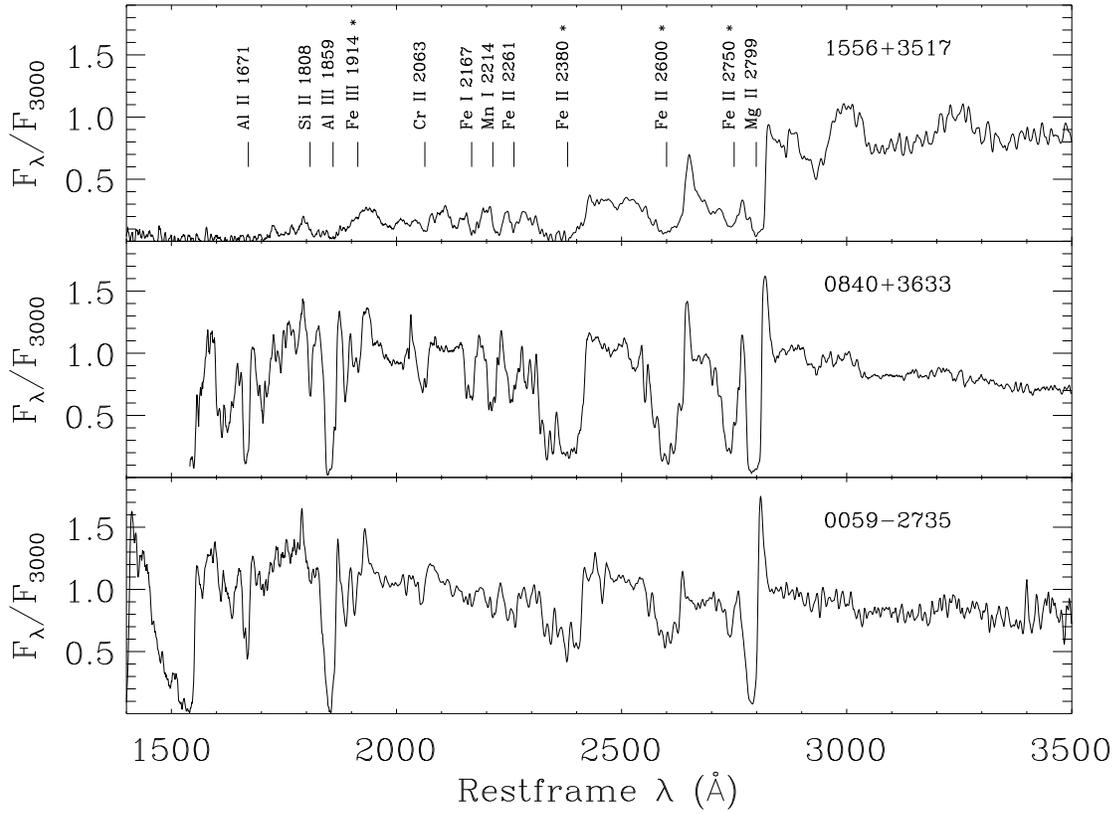}
\caption[figure1.ps]{Lick Kast 3m spectra of 0840+3633 and
1556+3517 compared to the spectrum of 0059-2715 from Weymann et al.\
(1991).  Prominent absorption features are marked; an asterisk denotes
metastable states of iron.}
\end{figure}

\end{document}